\documentclass[prb,twocolumn,showpacs,eqsecnum,amsmath,amssymb,floatfix,superscriptaddress]{revtex4-1}
\usepackage{amsmath}
\usepackage{graphicx} 
\usepackage{dcolumn} 
\usepackage{bm} 
\usepackage{hyperref} 
\usepackage{slashed}
\usepackage{subfigure}
\begin{document}

\title{Stripe melting, a transition between weak and strong symmetry protected topological phases}

\author{Yizhi You}
\affiliation{Department of Physics and Institute for Condensed Matter Theory, University of Illinois at Urbana-Champaign, 
Illinois 61801}
\affiliation{Kavli Institute for Theoretical Physics, University of California Santa Barbara, Santa Barbara, California 93106}
\author{Yi-Zhuang You}
\affiliation{Kavli Institute for Theoretical Physics, University of California Santa Barbara, Santa Barbara, California 93106}

\date{\today}
\begin{abstract}
For a gapped disordered many-body system with both internal and translation symmetry, one can define the corresponding weak and strong Symmetry Protected Topological (SPT) phases. A strong SPT phase is protected by the internal symmetry $G$ only while a weak SPT phase, fabricated by alignment of strong SPT state in a lower dimension, requires additional discrete translation symmetry protection. In this paper, we construct a phase transition between weak and strong SPT phase in strongly interacting boson system. The starting point of our construction is the superconducting Dirac fermions with pair density wave(PDW) order in $2d$. We first demonstrate that the nodal line of the PDW contains a $1d$ boson SPT phase. We further show that melting the PDW stripe and condensing the nodal line provoke the transition from weak to strong SPT phase in $2d$. The phase transition theory contains an O(4) non-linear-$\sigma$-model(NL$\sigma$M) with topological $\Theta$-term emerging from the proliferation of domain walls bound to an SPT chain. Similar scheme also applies to weak-strong SPT transition in other dimensions and predicts possible phase transition from $2d$ to $3d$ topological order.

\end{abstract}

\maketitle

\section{Introduction and Motivation}
Recently years, lots of attention had been devoted to the classification and characterization of symmetry protected topological(SPT) phases \cite{bi2015classification,metlitski2014interaction,xu2013wave,you2014symmetry,chen2012symmetry}.  An SPT state is short-range entangled(SRE) in the bulk, but with nontrivial boundary spectrum. As long as certain symmetry is preserved, one can never adiabatically connect the SPT state to a trivial many body state($e.g.$ direct product state). 

In particular, for many-body system with an internal symmetry, one can always define a corresponding weak and strong SPT phase\cite{fu2007topological,hughes2010Majorana,ramamurthy2014patterns,cheng2015translational}. A strong SPT phase in $(n+1)d$ is only protected by an internal symmetry $G$.  Meanwhile, a weak SPT phase in $(n+1)d$ can be constructed by alignment of strong SPT state in $(n)d$ protected by an internal symmetry $G'$ ($G'$ and $G$ can be different internal symmetries). Such corresponding weak SPT phase is therefore protected by both the internal symmetry $G'$ and the discrete translation symmetry along the alignment direction. One of the well-known example is the weak and strong time-reversal($\mathcal{T}$) invariant topological insulator(TI) in $3d$\cite{fu2007topological}. A strong TI in $3d$ can not be adiabatically connected to a trivial band insulator as long as  $\mathcal{T}$ is preserved. While a weak TI in $3d$, obtained by layer stacking of $2d$ $\mathcal{T}$ invariant TI along $z$ direction, requires both $\mathcal{T}$ and discrete translation symmetry along $z$ direction to protect the phase. Otherwise, without translation symmetry, one can always turn on strong tunneling between two TI layers and the fermion band becomes trivial without gap closing in the bulk. Therefore, as long as $\mathcal{T}$ is imposed, weak and strong TI belong to different phases and a connection between them requires gap closing.

Beyond the non-interacting fermion SPT phase described by band theory, much effort had been made on the exploration of strong/weak boson SPT phase with strong interaction. One representative of weak boson SPT phase in $2d$ can be constructed by alignment of spin $1$ Haldane chain along $y$ direction\cite{pollmann2012symmetry,haldane1983nonlinear,bi2015classification}. Such weak SPT state is protected by both time-reversal($\mathcal{T}$) and translation symmetry in $y$ direction. 

So far, the phase transition and the connection between weak to strong SPT phase at interacting level is less explored. In particular, as weak SPT phase is built upon the strong SPT state in a lower dimension, figuring out the connection between strong and weak SPT state also provides us a new way to relate SPT phases in different dimensions.

In this paper, we are going to answer the following question. Is there a straightforward way to connect weak and strong boson SPT states? In particular, how can we characterize the phase transition theory between strong and weak SPT phases?  The strategy we apply for these question is to melt the superconducting stripe.\cite{gopalakrishnan2013disclination,berg2009charge,agterberg2008dislocations,cho2014topological,mross2015spin} 

We start from $2d$ pair density wave superconducting state whose nodal line contains 8 copies of helical Majorana modes. These gapless modes can be gapped into a boson SPT phase, being equivalent to a spin 1 Haldane chain\cite{xu2013nonperturbative,you2014symmetry,bi2014bridging}. The pair density wave(PDW) state, as a stripe superconducting phase, contains alignment of nodal lines decorated with the Haldane chain which therefore forms a weak SPT state. By stripe melting and nodal loop condensation, the system becomes a $2d$ strong SPT state protected by $\mathcal{T}\times Z_2$ symmetry.  The similar construction also applies to $3d$ stripe superconductor where a $2d$ SPT phase lives inside the nodal plane of PDW order. The condensation of nodal membrane induces a transition from weak to strong SPT phase in $3d$.

In addition, after gauging the symmetry of an SPT state, the gauge flux can acquire nontrivial braiding statistics\cite{levin2012braiding,bi2014anyon}. In this paper, we would show that the connection between weak and strong SPT phases by stripe melting also provides a new way to generate the relation between intrinsic topological order in different dimensions.

Our stripe melting approach inherits the decorated domain wall idea raised by Chen $et$ $al.$\cite{chen2014symmetry}. In the decorated domain wall approach, they attach a lower dimension SPT phase on the domain wall of an Ising variable and the condensation of domain walls drives the system into a boson SPT state. In this work, we intend to find a natural and microscopic way to let the SPT state emerge inside the domain wall. The PDW stripes contain nodal line which act as a domain wall and the gapless fermion modes inside domain wall can be driven into a boson SPT state by interaction\cite{you2014symmetry,bi2014bridging,abanov2000theta}. Indeed, the phase transition between the stripe phase to the disordered phase includes a topological $\Theta$ term which reflects the nature of SPT decorated nodal line. Such decorated domain wall approach can be extended to other decorated defects, $e.g.$ the decorated vortex lines\cite{xu2015bosonic,tsui2015quantum,potter2015protection,bi2015construction,wang2015field}. In section \ref{Abrikosov}, we start from the Abrikosov phase of $3d$ topological superconductor whose vortex line contains $1d$ SPT chain and the melting of vortex lattice give rise to a boson SPT phase in $3d$.

\section{Melting the superconducting stripe in PDW phase}
In this section, we systematically investigate the phase transition from weak to strong SPT phase in all dimensions. 
The weak SPT state we start with throughout the paper is embedded in a stripe superconducting phase. The domain wall between different stripes contains a strong SPT state in a lower dimension which therefore makes stripe phase identical to a weak SPT state. After stripe melting and domain wall proliferation, the system experiences a transition from weak to strong SPT phase.

\subsection{stripe melting in 2d PDW}\label{2dpdw}
The starting point of our construction is the superconductivity in massless Dirac fermion at $2d$. Imagine we have a semimetal whose low energy theory contains several Dirac cones,
\begin{align} 
&H= \Psi^{\dagger}_{\bm{k}}(\sigma_x k_x+\sigma_z k_y)   \Psi_{\bm{-k}}
\end{align}
Here $\sigma$ acts on the spin index of the Dirac fermion.
Now we turn on $s$-wave pairing $\Delta=\Psi^{\dagger}_{\uparrow}\Psi^{\dagger}_{\downarrow}-\Psi^{\dagger}_{\downarrow}\Psi^{\dagger}_{\uparrow}$ of the Dirac cone\cite{fu2008superconducting}. The $s$-wave pairing term can emerge when we slightly doped the semimetal and turn on fermion interaction at finite strength, or from proximity effect between the Dirac fermion and an s-wave superconductor.
 The s-wave superconductivity turns the Dirac semi-metal into a gapped phase. Write the superconducting Dirac Hamiltonian in the Majorana basis $\Psi^{\dagger}_{\uparrow}=\chi_{1,\uparrow}+i \chi_{2,\uparrow}$,
\begin{align}
\label{pdw}
&H= \chi^T_{\bm{k}}(\sigma_x k_x+\sigma_z k_y+O_2\sigma_y \tau_x+O_1\sigma_y \tau_z)   \chi_{\bm{-k}} \nonumber\\
&\Delta=O_1+i O_2
\end{align}
Here $\tau$ matrix acts on the two Majorana index while $\sigma$ acting on spin index. Imagine we choose a specific gauge where $\Delta$ is real so $O_2$ is zero.

Now we turn to the situation that the superconducting field is nonuniform in space. It contains PDW order $\Delta=|\Delta| \cos(\bm{Qr})$ whose pairing amplitude modulates in space along the PDW wave vector and forms a stripy superconductor. Such PDW can be realized if the Dirac fermion dispersion has some nematicity and the fermi velocity is anisotropic. Else, the proximity effect between an s-wave PDW superconductor and the Dirac fermion can also induce the stripe superconducting Dirac cone.

The fermions in the stripes are gapped but the nodal line as a domain wall of $O_1$ contains $1d$ helical Majorana mode. Assume the PDW wave vector is in the $y$ direction so the nodal line is extended along $x$, the helical mode can be written as,
\begin{align} 
&H= \chi^T_{\bm{k}}(i \partial_x \sigma_x )   \chi_{\bm{-k}} \nonumber\\
&\chi=(\chi_{1,\uparrow},\chi_{2,\downarrow})
\end{align}
The only fermion bilinear mass $\chi^T\sigma_y \chi$ which can gap out this helical mode is associate with the imaginary part($O_2$) of the superconductivity in the stripe. As the nodal line has zero pairing strength, there is no way to gap out the helical mode.

Now assume we have 8 copies of the superconducting Dirac fermion embellished as Eq.\eqref{pdw}, the nodal line therefore contains eight helical Majorana modes. First we take the first 4 copies of the helical modes and couple them to an $O(3)$ vector $\vec{n}=(n_1,n_2,n_3)$,\cite{you2014symmetry,bi2014bridging,abanov2000theta}
\begin{align} 
\label{1d}
&H= \chi^T_{\bm{k}}(i \partial_x \sigma^{100} +n_1 \sigma^{312}+n_2 \sigma^{320}+n_3 \sigma^{332})   \chi_{\bm{-k}} \nonumber\\
&\mathcal{T}:\chi \rightarrow   \mathcal{K} \sigma^{300} \chi, ~~ \vec{n} \rightarrow   -\vec{n}
\end{align}  
$\sigma^{abc}$ refers to the direct product of Pauli matrices $\sigma^{a} \otimes \sigma^{b} \otimes   \sigma^{c} $. Here $\mathcal{T}$ symmetry operator is redefined in an unusual way as it has $\mathcal{T}^2=1$ when acting on the fermion sector. However, we would show in our later content that the $\mathcal{T}$ symmetry operator acts on the bosonic degree of freedom $\vec{n}$ in a nontrivial way.
As $\vec{n}$ becomes $-\vec{n}$ under $\mathcal{T}$ operation, any nonzero expectation value of $\vec{n}$ breaks $\mathcal{T}$. When $\vec{n}$  is disordered, we can integral out the fermion and obtain the effective theory of $\vec{n}$,
\begin{align} 
&\mathcal{L}= \frac{1}{g}(\partial_{\rho} \vec{n})^2+\frac{i\pi}{\Omega^2} \epsilon^{ijk}\epsilon^{\mu \nu} n^i \partial_{\mu}  n^j \partial_{\nu} n^k
\end{align} 
$\Omega^d$ refers to the volume of $S^d$.
The theory we obtain is the O(3) NL$\sigma$M with a topological theta term for $\Theta=\pi$, which is equivalent to a critical spin $1/2$ AF Heisenberg chain described by $SU(2)_1$ CFT. The theory is invariant under the $\mathcal{T}$ symmetry we defined in Eq.\eqref{1d}.  

Till now, we had coupled the first 4 copies of the helical Majorana modes with an $\vec{n}$ vector and the effective theory for $\vec{n}$ inherited from the fermion mode is a critical spin $1/2$ Heisenberg chain. 
The rest 4 copies of the helical mode can be coupled to another $\vec{n}'$ in the same way as Eq.\eqref{1d} which give rise to another O(3) NL$\sigma$M with a topological theta term for $\Theta=\pi$. The eight helical Majorana modes are therefore turned into two copies of spin $1/2$ Heisenberg chain described by $SU(2)_1$ CFT.

The two $SU(2)_1$ CFT can be fused into a gapped Haldane phase via marginal interaction as,\cite{white1996equivalence}
\begin{align} 
&\mathcal{L}= \frac{1}{g}[(\partial_{\rho} \vec{n})^2+(\partial_{\rho} \vec{n}')^2]+\frac{i\pi}{\Omega^2} \epsilon^{ijk}\epsilon^{\mu \nu} n^i \partial_{\mu}  n^j \partial_{\nu} n^k\nonumber\\
&+\frac{i\pi}{\Omega^2} \epsilon^{ijk}\epsilon^{\mu \nu} n'^i \partial_{\mu}  n'^j \partial_{\nu} n'^k+ \alpha~ \vec{n} \cdot  \vec{n}'
\end{align} 
When $\alpha$ is positive, the ferromagnetic interaction between two critical Heisenberg chain drives the theory into the Haldane phase, which is an SPT phase protected by $\mathcal{T}$ symmetry defined in Eq.\eqref{1d}.  As the interaction is marginal, we only need small interaction compared to the superconducting gap in the stripe so the stripe superconducting state is not affected. Finally, the eight helical Majorana modes in the nodal line are mapped into a Haldane chain\cite{white1996equivalence,shelton1996antiferromagnetic,kim2000phase} describe by NL$\sigma$M with a $\Theta=2\pi$ theta term,
\begin{align} 
\label{chain}
&\mathcal{L}= \frac{1}{g}(\partial_{\rho} \vec{n})^2+\frac{i2\pi}{\Omega^2} \epsilon^{ijk}\epsilon^{\mu \nu} n^i \partial_{\mu}  n^j \partial_{\nu} n^k
\end{align} 
The Haldane chain carries a free spin 1/2 zero mode on the boundary which is protected by $\mathcal{T}$. Based on the NL$\sigma$M description of the Haldane chain, the boundary of the $(1+1)d$ NL$\sigma$M in Eq.\eqref{chain} is a domain wall between $\Theta=2\pi$ to $\Theta=0$. The edge therefore contains an O(3) Wess-Zumino-Witten term at level one, with a two-fold degenerate ground state. We can represent the free spin 1/2 on the edge of the Haldane chain in terms of the $CP^1$ representation  as $\vec{n}=z^{\dagger}\vec{\sigma}z$. $z$ is a complex spinor field $z=(z_1,z_2)^T$ and $\mathcal{T}$ operator act on the spinor field as $\mathcal{T}: z \rightarrow   \mathcal{K} i\sigma^{2} z$. Therefore, the $\mathcal{T}$ symmetry acts projectively on the spinor $z$.

\begin{figure}[h]
    \centering
    \includegraphics[width=0.25
    \textwidth]{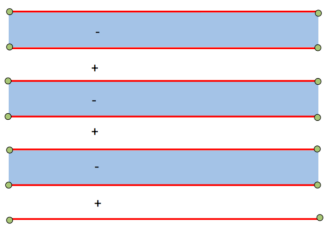}
    \caption{The stripe configuration of PDW. The pairing field in the white/blue stripes has opposite sign. The red line is the nodal line which contains a Haldane chain. The green dot is the free spin $1/2$ edge state of the Haldane chain.}
        \label{stripefig}
\end{figure}

The above construction decorates the nodal line with $1d$ SPT state, as is illustrated in Fig. \ref{stripefig}.

The PDW phase is therefore a weak SPT phase protected by both $\mathcal{T}$ and discrete translation symmetry in $y$ direction. The nodal line of the PDW order contains a Haldane chain as a $1d$ SPT phase protected by $\mathcal{T}$. Even the PDW state we start with is a fermonic theory, when the SC amplitude $\Delta$ and the O(3) vector magnitude is large enough, there would be an absence of fermonic excitation at low energy spectrum so the effective theory at low energy is bosonic. 

The discrete translation symmetry prevents two Haldane chain dimerizing into a spin 2 chain which is trivial under $\mathcal{T}$. Each end point of the nodal line contains a spin $1/2$ degree of freedom from the Haldane chain so the boundary along $y$ direction consists of an interacting spin 1/2 chain.  According to Lieb-Schultz-Mattis theorem\cite{lieb2004two,hastings2004lieb}, the GS of the spin 1/2 chain shall either be gapless or symmetry breaking (breaks $\mathcal{T}$ or translation).

Here and after, we would demonstrate that, after melting the superconducting stripe, there occurs a phase transition from the PDW phase to a uniform superconducting state, which meanwhile transmutes the weak SPT to a strong SPT state in $2d$. 

Melting the stripe superconducting configuration restores both translation and rotation symmetry. The melting procedure can be fulfilled by dislocation and disclination proliferation. The dislocation disorders the stripe configuration to restore the continuous translation symmetry while the disclination bends the straight nodal lines into an arbitrary crooked configuration. The stripe melting\cite{ryzhov1991dislocation,you2016geometry,cho2015condensation,gopalakrishnan2013disclination} driven by dislocation and disclination condensation would meanwhile condense the nodal loop. We assume there is some thermal or quantum fluctuation which effectively generates positive interaction between disclination/dislocations and therefore triggers a tendency to condense them. During the condensation, the origin Goldstone mode($\phi$) in the stripe phase associated with the translation symmetry broken is gapped by the vortex tunneling term $\cos(n \phi)$. The coherence of the condensed dislocations and disclinations generates a coherent state of all types of nodal loop configurations. Meanwhile, since the nodal line separates the positive and negative pairing amplitude, the nodal line must be closed in the bulk. If the nodal line has an end point in the bulk, there must be a half vortex of the pairing field associate with it. We focus on the situation where the system is vortex free so all the nodal line must form a close-loop configuration. 

After we proliferate and condense the nodal loops, the rotation symmetry of the pairing field is restored and one obtains a uniform phase. The ground state(GS) of the uniform phase can be written in terms of the superposition of all close nodal loop configurations decorated with Haldane chains.
\begin{figure}[h]
    \centering
    \includegraphics[width=0.35\textwidth]{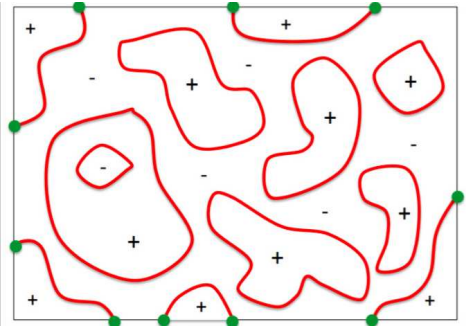}
    \caption{The GS wave function after nodal loop condensation. The system is saturated with all close nodal loops(red line) which carries a Haldane chain. The nodal loop separates positive/negative amplitude of the pairing field. The blue dots on the boundary are the free spin 1/2 edge states of the Haldane chain.}
        \label{nl}
\end{figure}

As each nodal line carries a Haldane chain, the GS wave function is a saturation of Haldane loop which separates the positive and negative pairing field. Let us assign the positive/negative pairing amplitude as a $Z_2$ variable and label it with $n_4$ so $\Delta=O_1=n_4$. In the PDW phase, $n_4=\cos(\bm{Qr})$  breaks $Z_2$ symmetry and forms a stripe configuration.  When the nodal loop condensed, the Ising scalar $n_4$ is then disordered and spatial symmetry is restored. The phase transition of such stripe order to disorder phase can be captured as,
\begin{align} 
&\mathcal{L}=  \kappa (\partial_{\mu} n_4)^2+\alpha (n_4-\bar{n}_4)^2 +\beta (n_4)^4+\sum^3_{a=1}\frac{1}{g}(\partial_{\mu} n_{a})^2\nonumber\\
&+\frac{i2\pi}{\Omega^3} \epsilon^{ijkl}\epsilon^{\mu \nu \rho} n^i \partial_{\mu}  n^j \partial_{\nu} n^k \partial_{\rho} n^l \nonumber\\
&\bar{n}_4=\cos(\bm{Qr})
\label{theta}
\end{align} 
The total degree of freedom we have is the $Z_2$ Ising variable $n_4$ between the nodal line and the O(3) vector $\vec{n}$ which comes from the NL$\sigma$M on the nodal line. The first three terms in Eq.\eqref{theta} are the common Landau-Ginzburg type theory for the phase transition. The fourth term is the fluctuation of the O(3) vector decorated on the nodal line. The last term is a topological theta term which captures the exotic physics during the transition. When the $Z_2$ Ising variable $n_4$ is ordered into stripe configuration, each nodal line $\partial_y n_4$ as a domain wall of the Ising variable carries a Haldane chain which could be described as a $O(3)$ NL$\sigma$M with theta term $\frac{2\pi}{\Omega^2} \epsilon^{ijk}\epsilon^{\mu \nu} n^i \partial_{\mu}  n^j \partial_{\nu} n^k$.  The last term in Eq.\eqref{theta} exactly captures the bound state between a $Z_2$ domain wall and the $(1+1)d$ theta term.  When $\alpha<0$, the $n_4$ scalar is disordered, the boson variable becomes $O(3) \times Z_2 \sim O(4)$, and the theory becomes the $(2+1)d$ O(4) NL$\sigma$M with $\Theta=2\pi$. This is known as an SPT phase protected by $\mathcal{T}\times Z_2$,
\begin{align} 
\mathcal{T}: n_{1,2,3} \rightarrow  -n_{1,2,3}; ~Z_2:  n_{1,2,3,4} \rightarrow  -n_{1,2,3,4}
\end{align} 
$\mathcal{T}$ acts on the O(3) vector on the nodal line and $Z_2$ symmetry enforces the $n_4$ Ising variable to be disordered so the nodal loops saturate and condense. The boson SPT phase with $\mathcal{T}\times Z_2$ protection has $(Z_2)^2$ classification, which contains two root states (with different symmetry transformation assignment) and each has a $Z_2$ classification\cite{bi2015classification,bi2014line}. Here since we uniquely assign the symmetry action, in disordered phase the classification of Eq.\eqref{theta} is $Z_2$. 

Here we emphasize that even the system we start with is fermionic, the finally SPT phase is a bosonic theory whose low energy effective theory can be absent from fermion excitations. This conclusion is based on the fact that the $O(4)$ vector magnitude which served as the mass of the fermion system is large enough so the fermion excitation is suppressed at low energy. In addition, one can also confine the fermonic degree of freedom by coupling the theory with a $Z_2$ gauge field\cite{you2014symmetry,bi2014bridging}. The vison flux of the $Z_2$ gauge field would trap 16 majorana zero modes which can be gapped out without breaking the $Z_2$ and $\mathcal{T}$ symmetry. Consequently, vison condensation would lead to a fully gapped nondegenerate bosonic state.

The effective theory we wrote in Eq.\eqref{theta} can also be verified in a microscopic way from the fermion side. We first take 4 out of the 8 superconducting Dirac cones and couple them with an O(3) vector\cite{you2014symmetry,bi2014bridging,abanov2000theta,fidkowski2011topological},
\begin{align} 
&H= \chi^T_{\bm{k}}(i \partial_x \sigma^{1000} +i \partial_y \sigma^{3000}+O_1 \sigma^{2300}+O_2 \sigma^{2100}\nonumber\\
&+n_1 \sigma^{2212}+n_2 \sigma^{2220}+n_3 \sigma^{2232})   \chi_{\bm{-k}} \nonumber\\
&\mathcal{T}:\chi \rightarrow \mathcal{K} \sigma^{2200}\chi , ~~Z_2: \chi  \rightarrow \sigma^{0100}\chi 
\label{2dspt}
\end{align}  

The first line illustrates the four copies of superconducting Dirac cone as Eq.\eqref{pdw}, the second line is the coupling between different copies of fermion via O(3) vector $\vec{n}$. The imaginary part of the pairing $O_2$ is taken to be zero by gauge choice so the pairing field is a real scalar field which can be regarded as an Ising variable $O_1$. The domain wall of $O_1$ contains four helical Majorana modes coupling with $\vec{n}$ as Eq. \ref{1d}.

The stripe melting and condensation of the nodal loops restored the translation(also rotation) symmetry and thus gapped the goldstone mode with respect to translation symmetry broken. The Ising variable $O_1$ is therefore disordered. Label $O_1$ as $n_4$, the four copies of Dirac cone now couples with an $ O(4)$ vector. When the O(4) vector is disordered, $\mathcal{T}\times Z_2$ symmetry is preserved. Integrating out the fermion give rise to a critical O(4) NL$\sigma$M with a theta term at $\Theta=\pi$.  

The rest 4 copies of the superconducting Dirac cones can be manipulated in the same way which gives rise to another critical $\Theta=\pi$ O(4) NL$\sigma$M. We can then fuse the two $\Theta=\pi$ theory into $\Theta=2\pi$ theory via ferromagnetic interaction between two O(4) vector,
\begin{align} 
\label{fuse}
&\mathcal{L}= \frac{1}{g}[(\partial_{\rho} \vec{n})^2+(\partial_{\rho} \vec{n}')^2]+\frac{i\pi}{\Omega^3} \epsilon^{ijkl}\epsilon^{\mu \nu \rho} n^i \partial_{\mu}  n^j \partial_{\nu} n^k \partial_{\rho} n^l\nonumber\\
&+\frac{i\pi}{\Omega^3} \epsilon^{ijkl}\epsilon^{\mu \nu \rho} n'^i \partial_{\mu}  n'^j \partial_{\nu} n'^k \partial_{\rho} n'^l+ \alpha~ \vec{n} \cdot  \vec{n}'
\end{align} 
Which finally drives the phase into a gapped SPT phase protected by $\mathcal{T} \times Z_2$ symmetry. (Here we point out that although we choose a specific gauge so the pairing field $\Delta=O_1$ is real, the phase fluctuation of the superconducting field would result in small fluctuation of $O_2$. However, as long as the ferromagnetic interaction between two O(4) vector is turned on, the theory is always in the gapped SPT phase even in the presence of small $O_2$. This would be demonstrated in the appendix.)
\begin{align} 
\label{O(4)}
&\mathcal{L}= \frac{1}{g}(\partial_{\rho} \vec{n})^2+\frac{i2\pi}{\Omega^3} \epsilon^{ijkl}\epsilon^{\mu \nu \rho} n^i \partial_{\mu}  n^j \partial_{\nu} n^k \partial_{\rho} n^l\nonumber\\
&\mathcal{T}: n_{1,2,3} \rightarrow  -n_{1,2,3}; ~Z_2:  n_{1,2,3,4} \rightarrow  -n_{1,2,3,4}
\end{align}
The edge state of this $\Theta=2\pi$ O(4) NL$\sigma$M is the O(4) Wess-Zumino-Witten theory at level one which is gapless. The GS wave function of this NL$\sigma$M can be written in terms of the partition function of an O(4) Wess-Zumino-Witten model\cite{xu2013wave}. 
\begin{align} 
|GS \rangle= \int D[\vec{n}] e^{i\frac{2\pi}{\Omega^{3}} \int d x^2 \int_0^1 du  \epsilon^{ijkl}\epsilon^{\mu \nu \rho} n_i \partial_{\mu}n_j \partial_{\nu} n_k\partial_{\rho}n_l} |\vec{n} \rangle
\end{align}
The wavefunction is a superposition of all possible configuration of $\vec{n}$, each configuration has a coefficient in terms of the Wess-Zumino-Witten term.
If we further break the O(4) variable to $Z_2$, the wave function is equivalent to the Levin-Gu model\cite{levin2012braiding}.

\subsection{Membrane condensation in 3+1D}\label{membrane}
In the previous section, we studied the PDW melting transition which connects a weak to strong SPT state in $2d$. Such idea can be easily generalized into other dimensions. In $3d$, we start with Weyl fermions and turn on s-wave pairing $\Delta=\Psi^{\dagger}_{\uparrow}\Psi^{\dagger}_{\downarrow}-\Psi^{\dagger}_{\downarrow}\Psi^{\dagger}_{\uparrow}$ to gap out the Weyl cone. The Hamiltonian of superconducting Weyl fermion can be written in Majorana basis as,
\begin{align} 
&H= \chi^T_{\bm{k}}(i \partial_x \sigma^{30} + i\partial_y \sigma^{10}+ i\partial_z \sigma^{22}+O_1 \sigma^{21}+O_2 \sigma^{23}) \chi_{\bm{-k}} \nonumber\\
&\mathcal{T}=\mathcal{K} i \sigma^{21}, ~~\Delta=O_1+i O_2
\end{align}  
The time reversal operator $\mathcal{T}$ we defined here is different from the usual one. The imaginary part of the pairing field $O_2$ is zero, by gauge choice.  Imagine the superconducting state has a PDW order where the amplitude of the pairing field modulates along $z$ direction as $\Delta=O_1=\cos(Q z)$. The superconductivity then forms a slab configuration along $z$ and $Q z=(n+1/2) \pi$ is the nodal plane which separates the positive and negative slab of the pairing field. In the nodal plane, the pairing strength is zero so the nodal plane contains a gapless Majorana cone,
\begin{align} 
&H= \chi^T_{\bm{k}}(i \partial_x \sigma^{3} + i\partial_y \sigma^{1})\chi_{\bm{-k}}
\end{align}
The only fermion bilinear mass $\chi^T \sigma^{2} \chi$ for the Majorana cone is associate with the imaginary part of the pairing field in the bulk, which is expected to be zero at the nodal plane. 
\begin{figure}[h]
    \centering
    \includegraphics[width=0.35\textwidth]{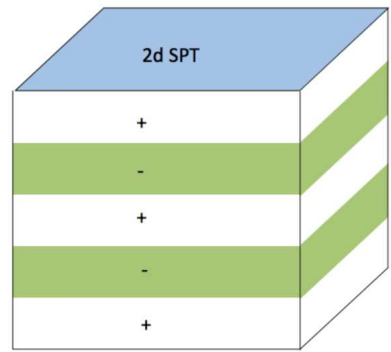}
    \caption{The slab configuration of PDW. the green/while slab contains pairing strength with positive/negative sign. The nodal plane(blue) contains a 2d SPT state.}
        \label{slab}
\end{figure}

Now we take 16 copies of such superconducting Weyl fermion with pair density wave order, each nodal plane then contains 16 Majorana cones.  

We first take 8 out of 16 Majorana cones and couple them to an O(4) vector in the same way as Eq.\eqref{2dspt}, 
\begin{align} 
\label{eightcone}
&H= \chi^T_{\bm{k}}(i \partial_x \sigma^{1000} +i \partial_y \sigma^{3000}+n_4 \sigma^{2100}\nonumber\\
&+n_1 \sigma^{2212}+n_2 \sigma^{2220}+n_3 \sigma^{2232})   \chi_{\bm{-k}} \nonumber\\
& Z_2:~  \chi \rightarrow  \sigma^{0300} \chi,~ \vec{n} \rightarrow  -\vec{n}
\end{align}  
The O(4) vector $\vec{n}$ change sign under the $Z_2$ symmetry we defined. When $\vec{n}$ is disordered, the theory is $Z_2$ invariant. Integrating out the fermions, the nodal plane then becomes a critical O(4) NL$\sigma$M with $\Theta=\pi$.  

The rest 8 Majorana cones couple to another O(4) vector in the same way as Eq.\eqref{eightcone}, we 
eventually obtain two copies of the critical O(4) NL$\sigma$M with $\Theta=\pi$. Couple the two critical theory of $\vec{n}$ through the ferromagnetic inter-copy interaction in the same way as we did in Eq. \ref{fuse}, the system goes into a gapped phase described by O(4) NL$\sigma$M with topological theta term at $\Theta=2\pi$. 
\begin{align} 
&\mathcal{L}= \frac{1}{g}(\partial_{\mu} n_{a})^2+\frac{i2\pi}{\Omega^3} \epsilon^{ijkl}\epsilon^{\mu \nu \rho} n^i \partial_{\mu}  n^j \partial_{\nu} n^k \partial_{\rho} n^l  \nonumber\\
& Z_2: \vec{n} \rightarrow  -\vec{n}
\end{align} 
This effective theory describes a $2d$ SPT state protected by $Z_2$ symmetry\cite{bi2015classification,you2014symmetry}.
Till now we had figured out that interaction can drive the 16 Majorana cones on the nodal plane into a gapped $2d$ boson SPT phase protected by $Z_2$ symmetry. In the PDW phase, a $2d$ boson SPT state living on the nodal plane stacks along the $z$ direction as Fig. \ref{slab}, which accordingly forms a weak SPT phase in $3d$ protected by $Z_2$ and discrete translation symmetry in $z$ direction.

Restoring the spatial symmetry of the PDW order would meanwhile drive the weak SPT to a strong SPT state. To demonstrate, first we melt the superconducting slabs. The melting procedure can be realized by dislocation and disclination proliferation, which bends the nodal plane into arbitrary close membrane configuration. The condensation of nodal membrane restores the rotation/translation symmetry. After slab melting, the GS wave function of the isotropic phase can be written in terms of the superposition of all possible close membrane configurations. If we relabel $O_1=n_5$ as an $Z_2$ variable, the nodal membrane condensation procedure is captured by the $Z_2$ order-disorder transition with a topological theta term,
\begin{align} 
\label{3d}
&\mathcal{L}=  \kappa (\partial_{\mu} n_5)^2+\alpha (n_5-\bar{n}_5)^2 +\beta (n_5)^4+\sum^4_{a=1}\frac{1}{g}(\partial_{\mu} n_{a})^2\nonumber\\
&+\frac{i2\pi}{\Omega^4} \epsilon^{ijklm}\epsilon^{\mu \nu \rho \lambda} n^i \partial_{\mu}  n^j \partial_{\nu} n^k \partial_{\rho}  n^l  \partial_{\lambda} n^m \nonumber\\
&\bar{n}_5=\cos(\bm{Qz})
\end{align} 
The first three terms are the Ginzburg-Landau theory of an Ising transition, the fourth term characterizes the fluctuation of the $O(4)$ vector $(n_1,n_2,n_3,n_4)$.
The last term indicates the domain wall of $n_5$ is bound to a topological $\Theta$ term in $2d$, as a signature of the $2d$ SPT state decorated in the nodal plane. 
When $n_5$ has stripe order, the nodal plane together with the $2d$ SPT situates along $z$ direction and simultaneously forms a weak SPT in $3d$.

At the disordered phase of $n_5$, $(n_1,n_2,n_3,n_4,n_5)$ together forms an $O(5)$ vector. The effective theory is just the O(5) NL$\sigma$M with $\Theta=2\pi$ topological theta term. This model is known as a boson SPT phase in $3d$ protected by $Z_2 \times \mathcal{T}$\cite{bi2015classification,you2014symmetry}. The symmetry operator acting on the O(5) vector is,
\begin{align} 
\mathcal{T}: n_{5} \rightarrow  -n_{5}; ~Z_2:  n_{1,2,3,4} \rightarrow  -n_{1,2,3,4}
\end{align} 

The effective theory in Eq.\eqref{3d} can also be obtained in a microscopic way from the fermion side. Take 8 out of 16 superconducting Weyl cones and coupled them with an O(4) vector\cite{you2014symmetry},
\begin{widetext} 
\begin{align} 
\label{3dbulk}
&H= \chi^T_{\bm{k}}(i \partial_x \sigma^{30000} + i\partial_y \sigma^{10000}+ i\partial_z \sigma^{22000}+O_1 \sigma^{21000}+n_1 \sigma^{23212}+n_2 \sigma^{23220}+n_3 \sigma^{23232}+n_4 \sigma^{23100} )\chi_{\bm{-k}} \nonumber\\
&\mathcal{T}: \chi \rightarrow \mathcal{K} i \sigma^{21000} \chi,n_{5} \rightarrow  -n_{5} ~~~~~ Z_2: \chi \rightarrow \sigma^{00300} \chi,  n_{1,2,3,4} \rightarrow  -n_{1,2,3,4}
\end{align}  
\end{widetext} 
$O_1$ is the real part of the pairing field(the imaginary part of the paring is zero by gauge choice). After nodal membrane condensation, the PDW becomes disordered($\langle \chi^T_{\bm{k}} \sigma^{21000}\chi_{\bm{-k}}\rangle=0$), the $\mathcal{T}\times Z_2$ symmetry we defined in Eq.\eqref{3dbulk} is therefore restored. Further label $O_1$ as $n_5$ and integrating out the fermion gives the critical O(5) NL$\sigma$M with $\Theta=\pi$, which is a gapless boson theory.

Couple the rest 8 superconducting Weyl cones with another O(4) vector in the same way and follow the same manipulation subsequently, we obtain another critical O(5) NL$\sigma$M with $\Theta=\pi$. Take the two copies of gapless O(5) NL$\sigma$M with $\Theta=\pi$ and turn on ferromagnetic inter-copy interaction as we did in Eq. \ref{fuse}, the system becomes a gapped phase described by O(5) NL$\sigma$M with $\Theta=2\pi$. (Here we point out that although we choose a specific gauge so the pairing field $\Delta=O_1$ is real. As long as the ferromagnetic interaction between two O(5) vector is turned on, the theory is always in the gapped SPT phase even in the presence of small $O_2$ fluctuation. This would be demonstrated in the appendix.)
\begin{align} 
\label{3dnlsm}
&\mathcal{L}= \sum^5_{a=1}\frac{1}{g}(\partial_{\mu} n_{a})^2+\frac{i2\pi}{\Omega^4} \epsilon^{ijklm}\epsilon^{\mu \nu \rho \lambda} n^i \partial_{\mu}  n^j \partial_{\nu} n^k \partial_{\rho}  n^l  \partial_{\lambda} n^m \nonumber\\
&\mathcal{T}: n_{5} \rightarrow  -n_{5}; ~Z_2:  n_{1,2,3,4} \rightarrow  -n_{1,2,3,4}
\end{align} 
This theory is a $3d$ SPT phase protected by $Z_2 \times \mathcal{T}$ symmetry. As we had emphasized in the last section, although the starting point of out model is a fermonic system, once the O(5) vector magnitude is large enough, the low energy spectrum is absent from fermion excitations and we can therefore view the effective theory as a bosonic one a low energy. 

In general, the $3d$ $Z_2 \times \mathcal{T}$ boson SPT state has $Z^3_2$ classification, which contains three root state with respect to different symmetry assignment and each root state has $Z_2$ class. Here we had specified the symmetry assignment so the classification is $Z_2$. The gapless surface state of this SPT phase is described by an O(5) Wess-Zumino-Witten model\cite{xu2013nonperturbative,yao2010topological}. If we further break the O(5) vector on the surface to $U(1) \times U(1)$ or $Z_2 \times Z_2$, the surface can be gapped and exhibit $Z_2$ topological order\cite{metlitski2014interaction,wang2014interacting}.

In summary, in this section we start from $3d$ PDW phase whose nodal plane contains a $2d$ SPT state. Melting the nodal plane leads a transition from weak to strong SPT state in $3d$. The strong SPT phase can be described by O(5) NL$\sigma$M with $\Theta=2\pi$. Such PDW melting construction for weak-strong SPT transition can be extended in other dimensions following the similar strategy. 

The strong SPT state we obtained via PDW melting is very similar to the decorated domain wall construction studied by several pioneers\cite{chen2014symmetry,xu2015bosonic,tsui2015quantum,potter2015protection,bi2015construction,wang2015field}. In our work, the domain wall is the nodal line(plane) of the PDW and the Ising variable is the positive/negative amplitude of the pairing field. As the PDW nodal line(plane) carries some gapless fermion mode, we can turn on interaction to drive the gapless fermion state into a gapped boson SPT state. In this way, the SPT state embellished on the domain wall appears in a nature way. This decorated defect picture can be extended to a large class of SPT states, some of which are beyond group cohomology classification\cite{xu2015bosonic,bi2015construction,wang2015field}.  In our PDW melting transition, the change of "topology" and the restoration of spatial symmetry appears at the same time as the PDW nodal line(plane) which breaks the continuous translation symmetry is embedded with O(n) $NL\sigma M$ with a topological $\Theta$ term. The condensation of the nodal line restore the spatial symmetry and meanwhile generates a topological $\Theta$ term from the emergent O(n+1) vector.

In particular, it was pointed out that the decorated domain wall approach only works when the SPT state living on the domain wall satisfied the Ònon-double-stacking conditionÓ (NDSC)\cite{tsui2015quantum,tsui2015topological}, otherwise one can always stacking a different counterpart to gap the boundary without degeneracy. Therefore, our stripe melting approach for weak-strong SPT phase transition is not general and is limited to the boson SPT states within the topological $\Theta$ term scheme\cite{xu2015bosonic}. In addition, even the strong SPT state does not require discrete translation symmetry protection, it demands additional $Z_2$ or $\mathcal{T}$ symmetry compared to the weak SPT phase in the stripe phase. The additional symmetry is essential as it ensures the $Z_2$ variable between the nodal line(plane) is disordered so the nodal line(plane) is condensed at the strong SPT phase. Accordingly, our strong SPT phase after stripe melting is protected by an enlarged (internal)symmetry which ensures the domain wall proliferation.

\section{line defect condensation in nematic paramagnetic}
Besides the stripe melting of PDW in Dirac fermions, here we propose another feasible realization on the phase transition from weak to strong SPT state in frustrated spin systems. Our starting point is a spin nematic paramagnetic state whose line defect contains a $1d$ SPT chain. When the line defects align in parallel into stripe configuration, the system as a $2d$ weak SPT state contains alignment of $1d$ SPT chain. Once the line defects proliferate and condense, the effective theory is akin to a strong $2d$ SPT phase.

\begin{figure}[h]
    \centering
    \includegraphics[width=0.4\textwidth]{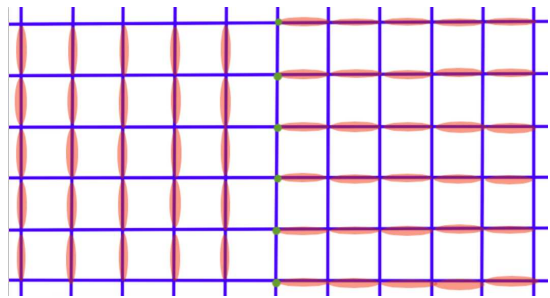}
    \caption{line defect of nematic paramagnetic state. Left and right side refers to opposite bond nematic order $v_1$. The red bond are the spin singlet bond in the AKLT chain. The line defect in the center carries a spin 1/2 chain.}
    \label{line}
\end{figure}

We first start from a spin $1$ nematic paramagnetic state on square lattice studied by a series of theoretical and numerical papers, \cite{wang2015nematicity,gong2015quantum}. 
\begin{align} 
 H=J_1 \sum_{\langle ij \rangle} \vec{S}_i \vec{S}_j+ J_2 \sum_{\langle \langle ij \rangle \rangle} \vec{S}_i \vec{S}_j+...
\end{align} 
The spin has antiferromagnetic interaction between nearest and next nearest sites. When $J_1/J_2$ is in the intermediate regime, some numerical evidence and theoretical prediction \cite{wang2015nematicity,gong2015quantum} argue that there exist a nematic paramagnetic phase which breaks spatial $C_4$ symmetry to $C_2$.
To verified this prediction, Wang $et.al.$\cite{wang2015nematicity} raised an exactly solvable Hamiltonian for the spin $1$ nematic paramagnetic state on square lattice,
\begin{align} 
&H=\sum_{ijk} P_3(\vec{S}_i+\vec{S}_j+\vec{S}_k)\nonumber\\
&P_3(S) = \frac{1}{720}S^2(S^2-2)(S^2-6) 
\end{align} 
The Hamiltonian is the sum of all projection operators and $P_3$ projects three spins onto $S=3$. The three spins $ijk$ live on the all elementary triangles where $ij$ and $jk$ are the NN bond and $ki$ are the NNN bond. The GS of each triangle shall contain at least one singlet bond to minimized the energy.  The GS of such Hamiltonian is an alignment of the AKLT chain\cite{affleck1987rigorous,wang2015nematicity} along either $x$ or $y$ direction (or $x+y, x-y$ direction) as illustrated in Fig. \ref{line}.  The GS has a bond nematic order which breaks $C_4$ rotation. The nematic order parameter can be written as a director field $v_1,v_2$,
\begin{align} 
&v_1=\frac{v^2_x -v^2_y}{\sqrt{v_x^2+v_y^2}}~,v_2= \frac{2 v_x v_y}{\sqrt{v_x^2+v_y^2}} \nonumber\\
&v_x=\vec{S}_{x,y} \vec{S}_{x+1,y}-\vec{S}_{x,y} \vec{S}_{x-1,y} \nonumber\\
&v_y=\vec{S}_{x,y} \vec{S}_{x,y+1}-\vec{S}_{x,y} \vec{S}_{x,y-1}
\end{align} 
The line defect of the nematic order, as the domain wall of $v_1$, consists a spin 1/2 chain with AF interaction. The line defect then carries a critical Heisenberg chain described by $SU(2)_1$ CFT illustrated in Fig. \ref{line}. 

Imagine we have two copies of such bond nematic paramagnetic state, the line defect contains two copies of critical spin 1/2 Heisenberg chain. Turn on a weak Ferromagnetic interaction between the two copy, the two $SU(2)_1$ CFT in the defect line can therefore be gapped and transmute into a spin 1 Haldane chain. Consequently, for two copies of such nematic paramagnetic phase, the line defect which separates opposite nematic order $v_1$ carries a $1d$ SPT state protected by $\mathcal{T}$.  When the bond nematic variable has stripe order $v_1= |v_1|cos(Qy)$, the defect line together with Haldane chain situates uniformly along $y$ direction and therefore produce a weak SPT phase in $2d$. Here we need to mentioned that when the line defect has a snake configuration going along the corners, the spins on defect line would also encounter with NNN antiferromagnetic interaction which have a tendency for dimerization. However, as long as the intercopy interaction is appropriate and strong enough, one can always drive the spins on the defect line into the Haldane phase\cite{white1996equivalence}.

Take the positive and negative value of $v_1$ as a $Z_2$ variable, once we proliferate the line defect and condense them, the defect line saturates in space and the $Z_2$ symmetry is therefore restored. The GS wave function then becomes a superposition of close defect loop configurations decorated with a $1d$ SPT.  This defect line condensation procedure in a nematic paramagnetic state is akin to the nodal line condensation in PDW we discussed in previous sections. At this stage, we obtain a strong SPT phase protected by $\mathcal{T}\times Z_2$ symmetry whose effective theory is the same as the condensed nodal line phase we explained in Eq.\eqref{2dspt}.

\section{Abrikosov lattice melting}
\label{Abrikosov}
In our previous discussion, we trigger the transition from weak to strong SPT by decorated domain wall condensation. Beyond the domain wall defect, one can also start from other defects like vortex or skyrmion embellished with an SPT state in the corresponded dimension. The condensation of vortex/skyrmion simultaneously drives the phase into a strong SPT state.

In this section, we demonstrate another way to realize weak-strong SPT transition by vortex condensation. We start from $1d$ SPT states decorated in the Abrikosov lattice. By vortex condensation and Abrikosov lattice melting, the weak SPT phase, originate from the $1d$ SPT chain hosted in the vortex lattice, transmute into a strong SPT state in $3d$.

We first start with two Weyl cones with opposite chirality and turn on the $s$-wave intra-cone pairing, the effective theory is
\begin{widetext} 
\begin{align} \label{vortexlattice}
&H= \chi^T_{\bm{k}}(i \partial_x \sigma^{103} +i \partial_y \sigma^{303}+i \partial_z \sigma^{223} + O_1 \sigma^{210}+O_2 \sigma^{230} ) \chi_{\bm{-k}}  \nonumber\\
&\Delta=O_1+i O_2, Z_2: \chi \rightarrow \sigma^{020} \chi
\end{align}
\end{widetext} 

The vortex line of the s-wave superconductor traps a gapless Majorana helical mode. Now imagine the superconductor is in the Abrikosov phase, where the vortex line of $\Delta=O_1+i O_2$ is pinned periodically along $x,y$ direction and accordingly forms an Abrikosov vortex lattice\cite{abrikosov2003type}. At each vortex line along $z$ direction, there is a helical Majorana mode. 

Now we duplicate 8 copies of such system. The 8 copies of helical Majorana modes inside the vortex line can be turned into Haldane chain as section \ref{2dpdw}. 
Thus, with 8 copies of superconducting Weyl cone pair in the Abrikosov phase, each vortex line of the vortex lattice carries a Haldane chain. The Abrikosov phase therefore forms a weak $3d$ SPT state protected by $\mathcal{T}$ and discrete translation symmetry in $x-y$ plane. Once we melt the Abrikosov lattice by proliferation and condensation of the superconducting vortex, the system concurrently undergoes a transition from weak to strong SPT state.  

As the vortex line carries Haldane chain described by an $O(3)$ topological $\Theta$ term, the vortex current $J^v_{\mu \nu} $ shall minimal couple with the $O(3)$ theta term as,
\begin{align} 
&J^v_{\mu \nu} B_{\mu \nu}\sim\frac{i2\pi}{\Omega^4} \epsilon^{abc}\epsilon^{\mu \nu \rho \lambda} O^1 \partial_{\mu}  O^2 \partial_{\nu} n^a \partial_{\rho}  n^b  \partial_{\lambda} n^c +..
..\nonumber\\
&J^v_{\mu \nu}=\epsilon^{ij}\epsilon^{\mu \nu \rho \lambda} \partial_{\rho}O_i\partial_{\lambda} O_j ,B_{\mu \nu}=\frac{2\pi}{\Omega^2}\epsilon^{abc}  n_a \partial_{\mu}n_b\partial_{\nu} n_c
\end{align} 
$\vec{n}$ is the O(3) vector degree of freedom coming from the Haldane chain in the vortex line. $B_{\mu \nu}$ is a two form gauge field associated with the O(3) $\Theta$ term.

When vortex condenses, the U(1) symmetry is restored. The O(5) boson field (emerged from $O(3) \times U(1)$) is described by the O(5) NL$\sigma$M as Eq.\eqref{3dnlsm}.  The effective theory after vortex condensation can also be verified by the microscopic fermion model we start with in Eq. \ref{vortexlattice}.

We have in total 8 copies of the superconducting Weyl cone pair, take 4 of them to couple to an O(3) vector.
\begin{widetext}
\begin{align} 
&H= \chi^T_{\bm{k}}(i \partial_x \sigma^{10300} +i \partial_y \sigma^{30300}+i \partial_z \sigma^{22300}+ O_1 \sigma^{21000}+O_2 \sigma^{23000} 
+n_1 \sigma^{02110}+n_2 \sigma^{02130}+n_3 \sigma^{02122}) \chi_{\bm{-k}} \nonumber\\
& \mathcal{T}:  \chi \rightarrow \mathcal{K} i \sigma^{02100} \chi , n_{1,2,3}\rightarrow -n_{1,2,3}, ~~~~Z_2:\chi \rightarrow\sigma^{02000} \chi,  O_i \rightarrow -O_i 
\end{align} 
\end{widetext}
When vortex condenses, the $Z_2\times \mathcal{T}$ symmetry is restored. Integrating out the fermion gives a critical O(5) NL$\sigma$M with $\Theta=\pi$. 
The rest 4 copies are treated similarly which gives another critical O(5) NL$\sigma$M with $\Theta=\pi$. These two critical boson theory can be merged into a gapped boson SPT phase in the same way as section \ref{membrane}.  Label $O_1,O_2$ as $n_4,n_5$, we finally have the effective theory as,
\begin{align} 
&\mathcal{L}= \sum^5_{a=1}\frac{1}{g}(\partial_{\mu} n_{a})^2+\frac{i2\pi}{\Omega^4} \epsilon^{ijklm}\epsilon^{\mu \nu \rho \lambda} n^i \partial_{\mu}  n^j \partial_{\nu} n^k \partial_{\rho}  n^l  \partial_{\lambda} n^m \nonumber\\
&\mathcal{T}: n_{1,2,3} \rightarrow  -n_{1,2,3}; ~Z_2:  n_{4,5} \rightarrow  -n_{4,5}
\end{align} 
This is another root state of $\mathcal{T}\times Z_2$ SPT phase\cite{bi2015classification} with different symmetry assignment compared to Eq.\eqref{3dnlsm}. The Abrikosov lattice melting procedure drives the theory from weak to strong SPT state via the proliferation of vortex line embellished by lower dimension SPT state. These decorated defect approach also shed light on the relation between boson SPT phases in different dimensions.

\section{Slab melting, a connection between 2d to 3d topological ordered after gauging the symmetry}
In the previous discussion, it turns out that a strong SPT state can be obtained by condensation of domain wall(or vortex) decorated with lower dimension SPT phase. The decorated domain wall picture reveals the connection between SPT state in different dimensions at interacting level. Then, it is natural to ask does such decorated domain wall picture also applies for true topological matter with intrinsic topological order?  In this paragraph, we would try to make a connection between topological order in different dimension by gauging the symmetry and melting the slab.

In section \ref{membrane}, we studied the phase transition between $3d$ weak and strong SPT phase. In the weak SPT side, the system is composed of a series of layers as domain walls separating different $Z_2$ variables in the slab. The domain wall itself contains a $2d$ SPT state. Such $2d$ SPT state, described by O(4) NL$\sigma$M, can also be protected by the $Z^a_2\times Z^b_2$ symmetry\cite{bi2015classification}.
\begin{align} 
\label{braiding}
&\mathcal{L}= \sum^4_{a=1}\frac{1}{g}(\partial_{\mu} n_{a})^2+\frac{i2\pi}{\Omega^3} \epsilon^{ijkl}\epsilon^{\mu \nu \rho } n^i \partial_{\mu}  n^j \partial_{\nu} n^k \partial_{\rho}  n^l   \nonumber\\
& Z^a_2:n_{1,2} \rightarrow - n_{1,2}, Z^b_2:n_{3,4} \rightarrow - n_{3,4},
\end{align} 
If we couple the O(4) vector to the $Z^a_2\times Z^b_2$  gauge field\cite{bi2014anyon,levin2012braiding}, the vison excitation of the gauge field would display nontrivial braiding statistics. Imagine we develop a pair of visons($\pi$ flux) of $Z^a_2$ and $Z^b_2$, each vision is bind with a half vortex of $(n_1,n_2)$(or $(n_3,n_4)$). The braiding between two vison excitations can be calculated via the braiding of the $\pi$ vortex of $(n_1,n_2)$ and the $\pi$ vortex of $(n_3,n_4)$. The $\Theta$ term in Eq.\eqref{braiding} indicates the braiding between these two vortices give rise to a phase of $\pi/2$.

Now we go back to the $3d$ system. When we are in the weak SPT phase, the system consists layers of the $2d$ SPT state along $z$ direction. When we gauge the $Z^a_2\times Z^b_2$ symmetry, each nodal plane exhibits $2d$ topological order. However, the vison loop of $Z^a_2$(or $Z^b_2)$, bound with a half vortex loop of $(n_1,n_2)$(or $(n_3,n_4))$, is contractible so there is no nontrivial loop statistics in $3d$\cite{levin2012braiding}.  

After we melt the slab and restore the $Z_2$ symmetry of $n_5$, the theory is a $3d$ SPT state protected by $Z^a_2\times Z^b_2\times Z^c_2$ symmetry,
\begin{align} 
\label{braiding2}
&\mathcal{L}= \sum^5_{a=1}\frac{1}{g}(\partial_{\mu} n_{a})^2+\frac{i2\pi}{\Omega^4} \epsilon^{ijklm}\epsilon^{\mu \nu \rho \lambda} n^i \partial_{\mu}  n^j \partial_{\nu} n^k \partial_{\rho}  n^l  \partial_{\lambda} n^m \nonumber\\
&Z^a_2:n_{1,2} \rightarrow - n_{1,2}, Z^b_2:n_{3,4} \rightarrow - n_{3,4},Z^c_2:n_{4,5} \rightarrow - n_{4,5}
\end{align} 
As we gauge the $Z^a_2\times Z^b_2\times Z^c_2$ symmetry, each vison of the $Z_2$ gauge is bind with a half vortex and the $\Theta$ term in Eq.\eqref{braiding2} indicates the three loop braiding procedure\cite{wang2014braiding,bi2014anyon,jian2014layer}. Namely, when we braiding the vison loop $Z^a_2$ with the vison loop $Z^b_2$, both penetrated by the vison loop $Z^c_2$, there accumulate a Berry phase of $\pi/2$ as a signature of $3d$ topological order. 

This method provides us a new insight to connect the topological order in different dimension. As one obtains topological order after gauging the symmetry of an SPT state, we expect the $3d$ topological order\cite{jian2014layer,wang2014braiding} with nontrivial three loop statistics(or loop-particle statistics) can be obtained via condensation of domain wall decorated with a $2d$ topological order.

\section{WZW theory as a muti-critical, deconfined quantum criticaity}
The phase transition from a boson SPT state to a trivial state is always beyond Ginzburg-Landau-Wilson(LGW) paradigm. Just as the $2d$ boson SPT state in Eq.\eqref{O(4)}, described by the O(4) NL$\sigma$M with topological theta term at $\Theta=2\pi$, has the same symmetry as the trivial O(4) disordered phase where $\Theta=0$. The critical theory connecting the SPT phase with $\Theta=2\pi$ to a trivial phase with $\Theta=0$ involves a critical theory of NL$\sigma$M with $O(5)$ Wess-Zumino-Witten term\cite{xu2013nonperturbative,chen2013critical}. Meanwhile, the NL$\sigma$M in $2d$ with $O(5)$ Wess-Zumino-Witten term can also appear as the deconfined quantum critical point which connects the two phases with different symmetry breaking.  

\begin{figure}[h] 
    \centering
    \includegraphics[width=0.5\textwidth]{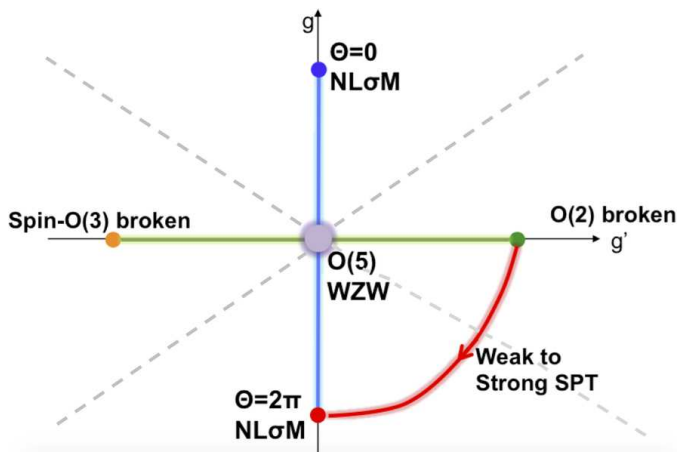}
    \caption{phase diagram}
    \label{all}
\end{figure}

To map these phenomenons in one unified diagram, we treat the NL$\sigma$M with $O(5)$ Wess-Zumino-Witten term as a multicritical point in Fig. \ref{all}. 

When we go from the bottom to top point, there happens a phase transition from the SPT phase($\Theta=2\pi$) to a trivial phase($\Theta=0$) illustrated in the blue line in Fig.\ref{all}. The center purple dot is the transition criticality represented by a NL$\sigma$M with $O(5)$ Wess-Zumino-Witten term.
Meanwhile, if we go from the left to right point(illustrated in the green line in Fig.\ref{all}), there happens a transition between two phases with different symmetry breaking\cite{senthil2004deconfined,vishwanath2013physics,moon2012skyrmions,grover2008topological}. Based on the $2d$ PDW system we discussed in section \ref{2dpdw}, the total symmetry we have is $O(3)\times O(2)$. O(3) degree of freedom comes from the Haldane chain we embellished in the nodal lines, while O(2) is the real and imaginary part of the superconducting pairing field. In the PDW phase(illustrated as the green dot in Fig.\ref{all}), the charge $U(1)$ symmetry is broken. The green line in Fig.\ref{all}  shows a transition 
from the O(3) spin rotation symmetry breaking state($e.g.$ AF order) to a PDW superconducting state, whose criticality is also described the $O(5)$ Wess-Zumino-Witten term.  

The PDW phase itself is also a weak SPT phase as it contains alignment of nodal line decorated with Haldane chain. After we melt the stripe and condense the nodal lines, the $Z_2$ symmetry is restored and we go into the strong SPT phase($\Theta=2\pi$) illustrated in the bottom red dot. The red line in Fig.\ref{all} displays the transition from weak to strong SPT phase. This phase transition is within the conventional LGW type which connects a symmetry broken state to an isotropic state. The isotropic state can be obtained via condensation of line defects($e.g$ nodal loop) hosting a topological $\Theta$ term in lower dimension, which therefore induce an SPT phase.

As a result, we had consistently unified three types of novel phase transition in a simple diagram.

\section{conclusion and outlook}

In this paper, we studied the transition from weak to strong SPT phases. For all the examples considered in this paper, we embellished the domain wall between different stripes with a topological theta term in lower dimension. The stripe melting procedure induced by domain wall saturation drives the system from a weak to strong SPT phase. This method can be view as a microscopic realization of the decorated domain wall(defect) approach\cite{chen2014symmetry,xu2015bosonic} as the SPT state embedded inside the nodal line(vortex line) can emerge from the gapless fermion mode via interactions. Our idea also cast the relation between topological superconductivity and boson SPT phases\cite{xu2015bosonic}. While the defects($e.g.$ nodal lines, vortex lines) inside the topological superconductor can carry multiple helical fermion modes, one can always gap out the fermion modes inside the defect into a gapped boson SPT state, and the condensation of the defects therefore generates a boson SPT phase. We hope this stripe melting idea could drive forward 
the exploration of SPT state from a microscopic point of view.

Further, we also suggested possible transition from $2d$ to $3d$ topological order via condensation of nodal planes decorated with topological order. We hope this idea would cast light on the exploration of $3d$ topological matter in condensed matter system.

In particular, we also discussed the relation between the three types of transition: an SPT to a trivial phase transition, a weak SPT to strong SPT phase transition and the transition between two ordered state with different symmetry broken.  We demonstrated that these three type of transition can be concluded in a unified diagram.

\begin{acknowledgements}
We are grateful to Thomas Scaffidi, Cenke Xu, Eduardo Fradkin, Andreas Ludwig, Steve Kivelson, Tarun Grover and Zhen Bi for insightful comments and discussions. This work was supported in part by the National Science Foundation through grants DMR-1408713 (YY) at the University of Illinois, grants NSF PHY11-25915(YY,YZY) at The Kavli Institute for Theoretical Physics, David and Lucile Packard Foundation(YZY) and NSF Grant No. DMR-1151208  (YZY). YY achokowledges the KITP Graduate fellowship program, where this work was initiated.
\end{acknowledgements}

\appendix
\section{SPT phase in the presence of $O_2$ fluctuation}
In section \ref{2dpdw} and \ref{membrane}, we choose a specific gauge so the pairing field $\Delta=O_1+iO_2$ is real and thus $O_2=0$. However, the phase fluctuation of the superconducting field would result in small fluctuation of $O_2$. Here we would demonstrated that as long as the ferromagnetic interaction between two O(4) vector is turned on, the theory is always in the gapped SPT phase even in the presence of small $O_2$. 

\begin{figure}[h] 
    \centering
    \includegraphics[width=0.45\textwidth]{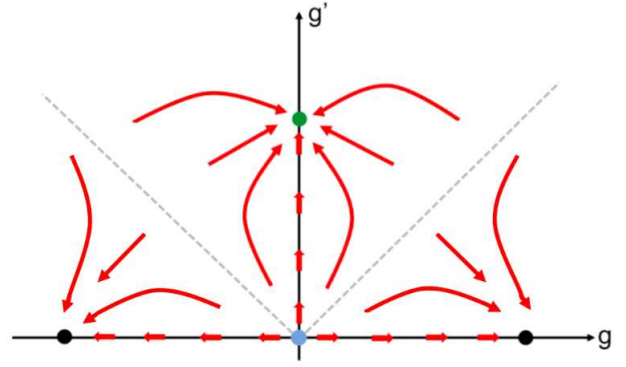}
    \caption{RG flow chart. $g$ is the strength of $O_2$ and $g'$ is the inter-copy coupling strength. The red arrows are the RG flow and the grey line is the phase boundary.}
    \label{app}
\end{figure}

In section \ref{membrane}, 8 copies of Weyl fermions with PDW order couples with $O(4)$ vector as,
\begin{widetext} 
\begin{align} 
\label{3dbulk}
&H= \chi^T_{\bm{k}}(i \partial_x \sigma^{30000} + i\partial_y \sigma^{10000}+ i\partial_z \sigma^{22000}+O_1 \sigma^{21000}+O_2 \sigma^{23000}+n_1 \sigma^{23212}+n_2 \sigma^{23220}+n_3 \sigma^{23232}+n_4 \sigma^{23100} )\chi_{\bm{-k}} 
\end{align}  
\end{widetext} 
If we restrict $O_2=0$, after membrane condensation which disorders $O_1$, the effective theory of $\vec{n}=(n_1,n_2,n_3,n_4,O_1)$ contains a topological theta term at $\Theta=\pi$. This describes a critical boson theory, illustrated as the blue dot in Fig.\ref{app}. 

As the phase of the superconducting field fluctuates, the nonzero value of $O_2$ could make the critical theory flowing into a gapped phase, illustrated as the black dots in the horizontal axis in Fig.\ref{app}. The red arrow in the horizontal line shows the direction of RG flow and $g$ is the strength of $O_2$ . 

When we have two copies of such system and turn on infinitesimal inter-copy Ferromagnetic interaction, if $O_2=0$, the two critical theory would be driven into a gapped SPT state($\Theta=2\pi$ ), illustrated as the green dot in the vertical axis in Fig.\ref{app}. 

If $O_2$ is small but nonzero, as long as the inter-copy Ferromagnetic interaction is large enough(above the grey phase boundary in Fig.\ref{app}), the theory would always flow to the gapped SPT phase.
This argument also works for the case in section \ref{2dpdw}.

\providecommand{\noopsort}[1]{}\providecommand{\singleletter}[1]{#1}%

\end{document}